\documentclass{article}
\usepackage[a4paper]{geometry}
\usepackage{graphicx}
\usepackage{amsmath}
\usepackage{upgreek}
\usepackage[english]{babel}
\usepackage{authblk}
\usepackage[separate-uncertainty=true]{siunitx}
\usepackage{enumitem}

\title{Pyroelectric Influence on Lithium Niobate During the Thermal Transition for Cryogenic Integrated Photonics}
\author[1]{Frederik Thiele}
\author[1]{Thomas Hummel} 
\author[1]{Nina Amelie Lange} 
\author[1]{Felix Dreher}
\author[1]{Maximilian Protte}
\author[1]{Felix vom Bruch}
\author[1]{Sebastian Lengeling}
\author[1]{Harald Herrmann}
\author[1]{Christof Eigner}
\author[1]{Christine Silberhorn}
\author[1]{Tim J. Bartley}
\affil[1]{Department of Physics \& Institute for Photonic Quantum Systems, Paderborn University, Warburger Str. 100, Paderborn, Germany}

\begin{document}
\maketitle

\begin{abstract}

Lithium niobate has emerged as a promising platform for integrated quantum optics, enabling efficient generation, manipulation, and detection of quantum states of light.
However, integrating single-photon detectors requires cryogenic operating temperatures, since the best performing detectors are based on narrow superconducting wires. While previous studies have demonstrated the operation of quantum light sources and electro-optic modulators in LiNbO$_3$ at cryogenic temperatures, the thermal transition between room temperature and cryogenic conditions introduces additional effects that can significantly influence device performance. In this paper, we investigate the generation of pyroelectric charges and their impact on the optical properties of lithium niobate waveguides when changing from room temperature to \SI{25}{K}, and vice versa. 
We measure the generated pyroelectric charge flow and correlate this with fast changes in the birefringence acquired through the Sénarmont method.
Both electrical and optical influence of the pyroelectric effect occurs predominantly at temperatures above \SI{100}{K}.
\end{abstract}

\section{Introduction}
The aim of integrated quantum photonics is to combine multiple optical devices on a single chip to profit from their interplay in a stable and scalable experimental setup~\cite{o2009photonic,pelucchi2022potential}. 
Components for quantum light generation, manipulation, and detection can be combined to realize a multitude of complex integrated circuits for various applications~\cite{kim2020hybrid,moody20222022}. 
A careful selection of the integrated components and material platform is required since the devices must be compatible in their optical characteristics and operation conditions~\cite{wang2020integrated}. 
While most devices are optimized for room temperature operation, superconducting single-photon detectors and solid-state quantum emitters require cryogenic temperatures~\cite{elshaari2020hybrid}. 
When combining cryogenic integrated components, these devices dictate the operation temperature for integrated chips. 
To that end, it is indispensable to develop all-integrated components for the cryogenic temperature application.

Lithium niobate is a well established nonlinear platform for integrated quantum photonics. In recent years, this platform has shown great cryogenic compatibility by the realisation of integrated components for generation, manipulation, and detection of quantum states of light~\cite{YOSHIDA1992, Lomonte2020,Thiele2020,Sahu2022,Thiele2022a,Lange2022}. Cryogenic single-photon sources in lithium niobate are realized through the integration of quantum dots~\cite{Aghaeimeibodi2018}, and via spontaneous parametric down-conversion in quasi-phase-matched waveguides~\cite{Lange2022,Lange2023}. 
The first cryogenic electro-optic modulators were demonstrated in the 90s~\cite{YOSHIDA1992, Sahu2022, Morse1994b, mccammon1994fiber, McConaghy1996}.
Nowadays three main modulator schemes exist for cryogenic operation, such as phase-shifter, directional coupler, and polarization converter~\cite{Thiele2020,Thiele2022a}.
Integrated superconducting single-photon detectors are successfully shown for weakly confined waveguide structures~\cite{Tanner2012,hopker2021integrated,Hopker2019}, and as well on thin film lithium niobate~\cite{Lomonte2020,Colangelo20}.
Electro-optic modulators can be used to read out superconducting devices~\cite{Youssefi2020,thiele2023optical_arxiv}, and are integrated together with superconducting detectors~\cite{Lomonte2020}. In addition, nonlinear frequency conversion processes have been established at cryogenic temperatures, such as second harmonic generation in periodically poled waveguides~\cite{Bartnick2021}, and cryogenic transducers for terahertz generation by optical rectification~\cite{huang2013high,carbajo2015efficient,huang2015highly}

The listed integrated devices in lithium niobate have demonstrated their functionality when operated at a stable cryogenic temperature. However, in previous work~\cite{Thiele2020,Bartnick2021,bravina2004low} we observed unpredictable changes in the electrical and optical properties during the thermal transition between room temperature and cryogenic temperatures. These changes range from sudden jumps to slower shifts in the optical properties on different time scales. The cryogenic operation of nonlinear interactions and electro-optic modulation induces a shift of the operation characteristics such as the phasematched wavelength, when compared to room temperature applications~\cite{Lange2022,Thiele2020,Bartnick2021}. 
The variation can be attributed to thermal stress and pyroelectric charges which build up inside the ferroelectric lithium niobate crystal when the temperature changes~\cite{lang1974sourcebook,jachalke2017measure}. 
The unbound charge carriers induce localized electric fields, causing microscopic refractive index perturbations, due to the electro-optic effect~\cite{Weis1985}. In addition, the unbound carriers can flow through the sample and accumulate at the surface. Electrodes placed on top of the sample can collect the charges such that a macroscopic current can be measured with a sensitive current meter. When operating the chip under ambient conditions, charges at the surface can be neutralized by charges in the air. However, when operating lithium niobate in an evacuated cryostat chamber, they cannot be so easily neutralized.
As a consequence, pyroelectric charges can disturb the device performance during a temperature transition \cite{Thiele2020, Bartnick2021}, or even damage integrated structures, such as superconducting detectors~\cite{hopker2021integrated}. 

In this paper, we investigate the origin of perturbations, which appear in the thermal transition. 
Characterizing and understanding the temperature dependent changes are necessary to optimize photonic circuits and to chose the optimal parameters such as the bias voltage for electro-optical modulators for cryogenic operation. 
Moreover, refractive index perturbations can influence the operation wavelength for phase-matched processes such as nonlinear frequency conversion~\cite{Bartnick2021}. Low temperature investigations of the pyroelectric behavior of lithium niobate were performed at a stable temperature, as well as during a thermal transition~\cite{Thiele2020,Bartnick2021, bravina2004low, brownridge2003saturation, Bravina2007, Herzog2008}.
We report to the best of our knowledge the first simultaneous measurement of electric discharges inside an electro-optic modulator and variations in the optical throughput, when changing the operation temperature. 
We observe the discharges by measuring the current between electrodes on the surface of the modulator, and use the Sénarmont method to visualize localised changes in the refractive index, manifesting as phase shifts in the transmitted light~\cite{Parravicini2011}.
The device under test is cooled down from room temperature to cryogenic temperatures and warmed up again to obtain data for a whole cool-down and warm-up, covering a temperature range from \SI{25}{K} to \SI{285}{K}.

\section{Pyroelectric refractive index variations}
Lithium niobate is used for integrated photonic circuits due to its large nonlinear properties, wide transparency range, and electro-optic properties~\cite{Weis1985,Sharapova2017}. 
In this material the refractive index can be varied by applying a constant electric field $E_j^{DC}$, inducing a refractive index in a correlated $i$-direction~\cite{Weis1985}. 
Since this material is birefringent, the amplitude of the refractive index change $\Delta n_{i}$ depends on the direction of the DC-electric field following~\cite{Weis1985, Nye1957, Boyd2007},
\begin{equation}
  \Delta n_{i} \sim  r_{ij} E_j^{DC}\textrm{,}  
\end{equation}
where $r_{ij}$ is the electro-optic tensor for lithium niobate.
An electric field applied in the vertical orientation modulates the refractive index according to the electro-optic field tensor $r_{33}$, and an applied horizontal field by $r_{13}$~\cite{Weis1985, Kotitz1990}.

\begin{figure}[ht]
    \centering
    \includegraphics[width=7cm]{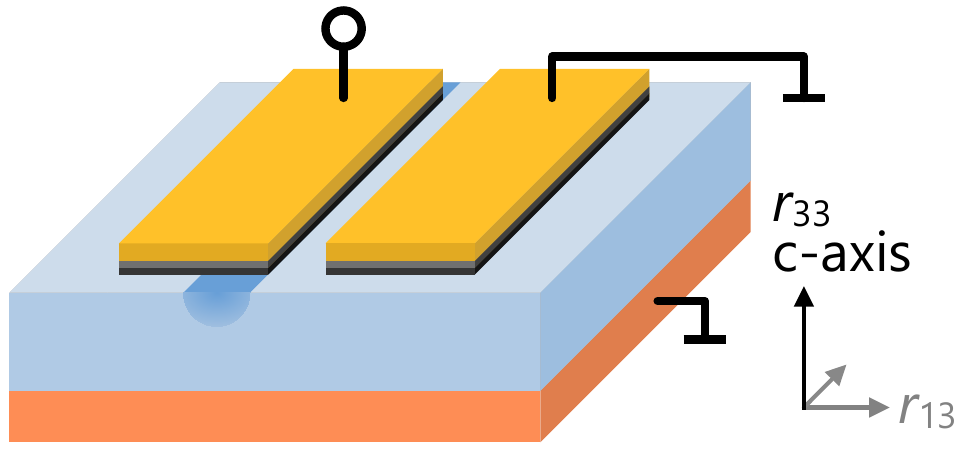}
    \caption{
    Schematic of the device under test to investigate electrical and optical changes in the thermal transition in a cryostat.
    The sample consist of a Ti-in-diffused z-cut lithium niobate waveguide with electrodes in a phase modulator configuration.
    The modulator is mounted on a copper plate, connecting the backplane of the sample to the case-ground of the cryostat.
    The electrodes are connected to an break out board via aluminium wirebonds and coaxial cables which transmits the signal to a room temperature current meter. 
    }
    \label{fig:PicModulator}
\end{figure}

We realise an electro-optic modulator in z-cut lithium niobate by fabricating titanium in-diffused waveguides with electrodes placed around the waveguide, see Fig.~\ref{fig:PicModulator}.
These electrodes allows us to apply the external electric field $E_j^{DC}$ to change the refractive index.
In this paper we use a phase modulator where one of the electrodes is placed on top of the waveguide to induce a vertical electric field, as depicted in Fig.~\ref{fig:PicModulator}.

In a thermal transition, lithium niobate generates unbound carriers in the material domain~\cite{Parravicini2011, Nye1957}, resulting in electric fields in the material due to these pyroelectric charges. 
The induced fields changes the refractive index independent of the applied voltage at the electrodes of the modulator. 
In z-cut lithium niobate, the unbound charge carriers can flow along the grain boundaries towards the top surface and recombine at the top surface~\cite{Sanna2017}. 
The redistribution and discharge of the carriers will follow a capacitive discharge with an associated resistance $R$ and capacity $C$~\cite{Korotky1996}
\begin{equation}
    I(t)= I_0 e^{-t/\tau_{RC}},
    \label{eq:decay}
\end{equation}
where $\tau_{RC}=RC$ is the time constant related to the discharge.
The charges accumulate in a confined space and therefore the resulting effect is a microscopic effect. 
In addition, we expect the pyroelectric charges to be generated at various locations in the lithium niobate.
The charge relaxation time, or discharge time constant $\tau_{RC}$ can thus be derived dependant on the microscopic geometry of the lithium niobate, and not on the macroscopic design of the electrodes.
Electrodes along the waveguide can collect the charge carriers and a connected current meter between two electrodes average the charge flow.
In a macroscopic perspective this is known as the pyroelectric effect which is described as a current between two domain surfaces while the temperature changes~\cite{Nye1957, Parravicini2011}.

\section{Methods}
We investigate the correlation between changes in the refractive index and generated pyroelectric discharges in lithium niobate while changing the operation temperature from room temperature to cryogenic temperatures and back to room temperature.
To measure changes in the refractive index, we use the birefringence. Light is coupled into orthogonal polarisation modes (TE and TM) of the waveguide and subsequently interfered, such that a change in the relative phase between two modes can be extracted. 
Simultaneously, the generated pyroelectric charges are measured as a current flowing through the electrodes on the surface of the sample. 
The simultaneous measurement of changes in the refractive index and generated charges may be correlated, since induced electric fields should result in refractive index changes. 

\subsection{Sénarmont method with electro-optic waveguides}
The Sénarmont-method measures the birefringence between TE and TM polarization modes of light in a waveguide~\cite{Parravicini2011}.
Both transmitted modes will accumulate a change in their optical path length and thus phase, due to the temperature dependent change in the refractive index~\cite{Jundt1997,Kotitz1990}. 
The relative phase between these modes is extracted by measuring the power ratio between the two polarization outputs after they are interfered.
Since this is a relative measure, it is independent of coupling losses.
Varying the temperature results in a change of optical coupling into the waveguide due to thermal contraction and expansion of the mounting stages.
However, crucially the Sénarmont-method is independent of the coupling efficiency.

\begin{figure}[ht]
    \centering
    \includegraphics[width=11cm]{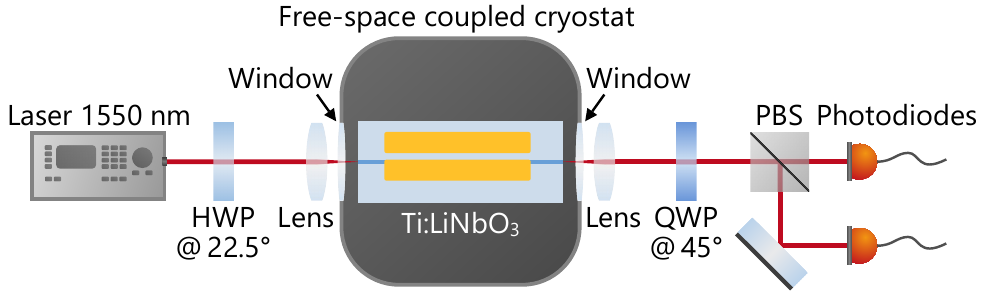} 
    \caption{
    Schematic of the setup to measure the optical changes in the lithium niobate sample during the thermal transitions. 
    The sample is mounted in a free-space accessible cryostat whereby light is coupled through the windows. Changes in birefringence in the waveguide are measured using the Sénarmont-method.
    The optical input comes from a \SI{1550}{nm} wavelength laser with \SI{2.2}{mW} of optical power.
    To measure the generated pyroelectric charges, the gold electrodes on the surface of the sample are wire bonded and connected to a current meter at room temperature.
    }
    \label{fig:PicSetup}
\end{figure}

We mount our lithium niobate waveguides in a free-space accessible cryostat and couple light into the waveguides, as it can be seen in Fig.~\ref{fig:PicSetup}.
The input light is polarized at \SI{45}{^{\circ}} such that light is coupled equally into the two polarizations.
After the transmission through the waveguide with the length $L$, a relative phase 
\begin{equation}
    \Phi_{\textrm{H,V}} = 2\pi n_{\textrm{H,V}}  L/ \lambda
\end{equation}
is accumulated for the horizontal and vertical polarization, where $n_{\textrm{H,V}}$ is the refractive index for either polarization, and $\lambda$ is the wavelength.

To acquire the relative phase difference $\Delta \Phi=\Phi_{\textrm{H}}-\Phi_{\textrm{V}}$, the two polarizations are interfered with a quarter-wave plate aligned at \SI{45}{^{\circ}} and a polarizing beam splitter (PBS).
The intensities after both output channels of the PBS $I_{1,2}$ are acquired and depend on $\Delta \Phi$ via 
\begin{equation}
   I_{1,2} = I_0 \cdot \frac{1}{2} \left(1 \pm \sin(\Delta \Phi)\right)~.
\end{equation}

In the experiment, the relative phase can be directly extracted from the intensity of one output of the PBS if the total in-coupled power $I_0$ is known. 
However, thermal contraction and expansion of the mounting stages changes the coupling efficiency resulting in a time-dependant $I_0$.
To extract changes in the refractive index, the intensity of one output arm of the PBS is normalized by the sum of both outputs
\begin{equation}
    \tilde{I}_{\textrm{1,2}} = \frac{I_{\textrm{1,2}}}{I_\textrm{0}} = \frac{I_{\textrm{1,2}}}{I_\textrm{1}+I_\textrm{2}}~,
    \label{eq:norm}
\end{equation}
which allows us to extract $\Delta \Phi$ independent of the coupling efficiency.

We change the sample temperature in the cryostat from \SI{280}{K} to \SI{25}{K} and back to \SI{280}{K} with a constant rate of \SI{1}{K/min}. 
The relative phase changes due to the temperature dependence of the refractive index. 
As a result, we expect to see an oscillation in the normalized output intensity with an oscillation time of \SI{3}-\SI{10}{min/oscillation}.
In addition, recombination of the pyroelectric charges can generate sudden changes in the refractive index due to the aforementioned electro-optic effect causing an abrupt change in the phase $\Delta \Phi$.

To measure the change in the generated pyroelectric charges we connect a current meter between the electrodes placed around the waveguide of the lithium niobate sample. 
The current meter has a sensitivity of \SI{0.05}{pA} when acquiring currents in a range up to $\pm\,$\SI{2}{nA}.
To correlate changes in the optical throughput and generated current, we measure both simultaneously with a sample rate of \SI{1.7}{samples/s}. 

Our input light comes from a CW-laser at a wavelength of \SI{1550}{nm} with \SI{2.2(1)}{mW} of optical power.
Lenses outside the cryostat couple the light into the waveguide.
At room temperature, the transmitted power through the waveguide is \SI{1.2(1)}{mW}.
The coupling efficiency changes during the thermal transition due to thermal contraction and expansion, dominated by height variations of the mounting stack.
For that reason, we reoptimize the vertical alignment of the sample during the thermal transition to ensure the transmitted power does not drop below \SI{6}{\micro W}. 
These brief optimizations allow the transmitted power to exceeded at least  \SI{300}{\micro W} after every reoptimization, while the measurement is interrupted for approximately only one minute for eight optimizations during the cooling process.

\subsection{Sample fabrication} \label{sec:fab} 
The device under test is a z-cut lithium niobate sample with a titanium in-diffused waveguide and electrodes in a phase-modulator configuration~\cite{Thiele2022a}.
To fabricate the waveguide, titanium is selectively deposited on the surface with a photo-lithography and lift-off process, resulting in \SI{5}{\micro m} wide titanium stripes. 
These stripes are in-diffused in the sample via heating in an oven, resulting in a localized change of the refractive index, allowing the guiding of light.
The linear losses of the waveguide are determined to be about \SI{0.15}{dB/cm}\cite{Thiele2022a} with a Fabry-Pérot method ~\cite{Regener1985}.
A second photo-lithography and lift-off process are used to fabricate electrodes on the sample.
The electrodes consist of a stack of \SI{400}{nm} silicon dioxide as a buffer layer, \SI{10}{nm} chromium for adhesion, and \SI{300}{nm} gold as our electrode.
The electrodes are placed in a phase modulator configuration to induce a primarily vertical electric field component in the waveguide. 
To do so, one electrode is placed on the top of the waveguide such that one inner electrode edge is aligned to the outer waveguide edge. The second electrode is placed at a distance of \SI{9}{\micro m} from the first electrode, as can be seen in Fig.~\ref{fig:PicModulator}.
The waveguides length is \SI{22}{mm} with \SI{12}{mm} long electrodes on the top surface. 
The samples total thickness of the lithium niobate sample is \SI{500}{\micro m}. 
Initial characterisations of the phase modulator with the Sénarmont method show a modulation voltage of \SI{23.3}{V} at room temperature and \SI{40}{V} at cryogenic temperatures.
The modulation voltage is the required voltage to switch from a maximum to a minimum in the intensity output.

\begin{figure}[ht]
    \centering
    \includegraphics{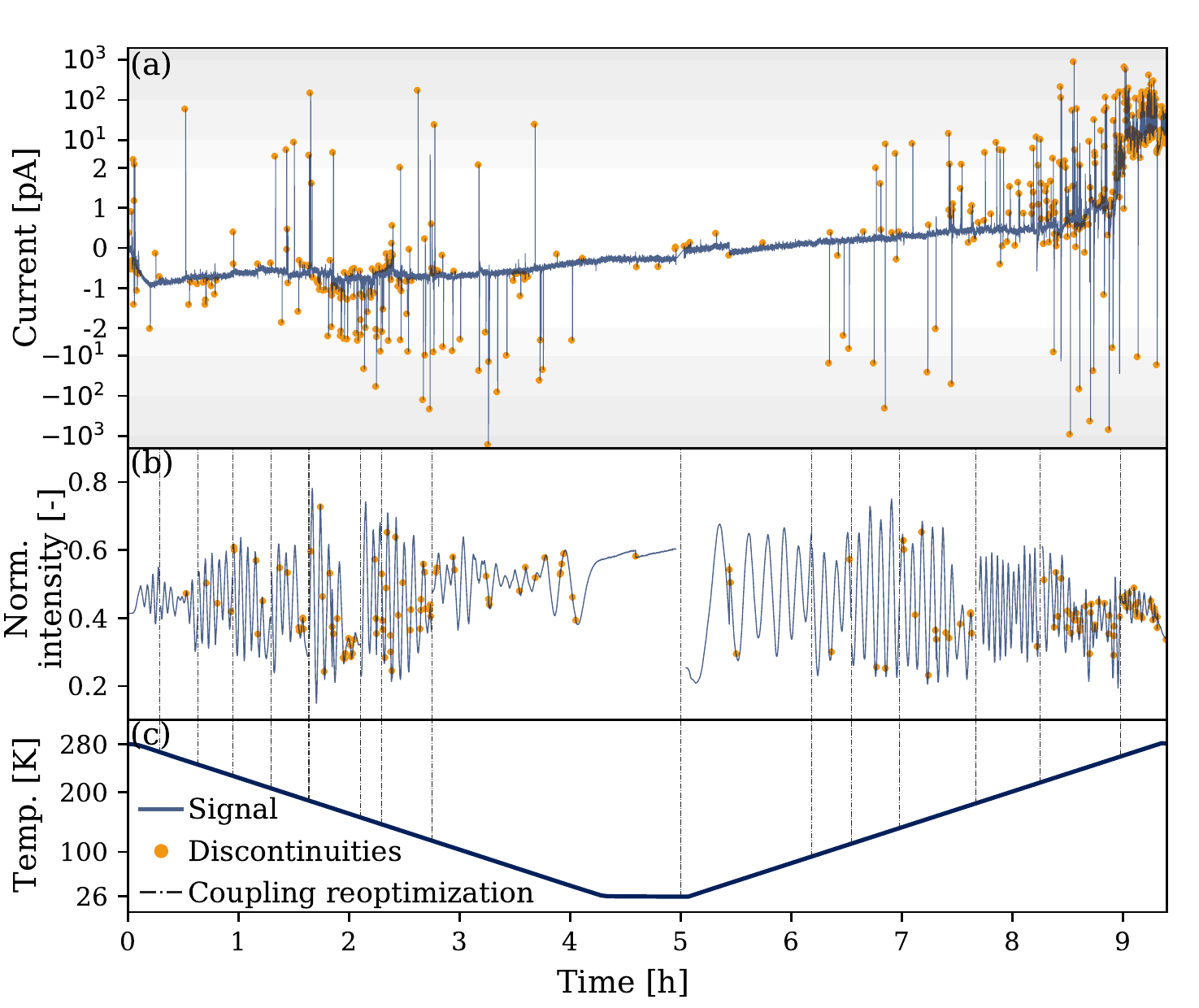}
    \caption{
    Measurement data of (a) the measured current between the electrodes of the phase modulator, (b) the normalized intensity $\tilde{I}_{\textrm{1}}$ in a single PBS port,  and (c) the temperature of the sample stage on which the sample is mounted.
    The acquired signals are represented in the blue lines, and every extracted discontinuity in the electrical or optical signal is marked with an orange dot.
    The vertical lines indicate the time periods where the optical power through the sample is reoptimized with the vertical alignment.}
    \label{fig:PicData}
\end{figure}

\section{Results}
The lithium niobate sample is cooled down and warmed up in the cryostat while the generated current and optical transmission are acquired.
These two signals and the temperature are shown in Fig.~\ref{fig:PicData}. 
In addition, the time intervals where the optical coupling is reoptimized are marked with a vertical dashed line.
In Fig.~\ref{fig:PicData}~(a) shows the corresponding electrical current measured through the electrodes. 
The normalized optical power of channel~$1$ is shown in Fig.~\ref{fig:PicData}~(b), and the expected \SI{3}- \SI{10}{min/oscillation} due to the thermal change of the refractive index can be seen.

Discontinuities in the optical and electrical signal can be extracted using a signal processing algorithm to minimize the human bias.
Our extraction method uses the SciPy package in our algorithm written in Python~\cite{Scipy2001}.
Since a discontinuity in the optical signal is a rapid change of the phase, we can extract this with a derivative.
We use an averaged derivative implemented as a convolution of the dataset with the matrix $\frac{1}{6}\left[1,1,1,-1,-1,-1\right]$.
The slow oscillation is then still present, but this is filtered out with a $4^{\mathrm{th}}$ order high pass Butterworth filter with a cut-off frequency $f_c=\SI{36}{mHz}$.
The discontinuities are then extracted with a the SciPy peakfinding algorithm with a prominence of $>0.006$, which indicates the minimum height of the peak compared to the base of the signal.
Fine-tuning of the analysis parameters was challenging because the height of the peaks varies in many orders of magnitude.
It was impossible to discriminate between optical changes and noise in the optical signal at constant cryogenic temperatures. 
For this reason, the optical discontinuities at constant cryogenic temperatures are selected by hand.
The electrical signal only requires a high-pass filter to remove the DC-drift current. After that, the discontinuities are extracted by the peakfinder algorithm.
We used a $4^{\mathrm{th}}$ order Butterworth filter with a cut-off frequency $f_c=$ \SI{0.27}{mHz}, and a minimum prominence of $0.2$ in the peak-finder.
All extracted discontinuities are shown as orange dots in Fig.~\ref{fig:PicData}~(a)~and~(b).

The displayed data range in Fig.~\ref{fig:PicData}~(a) highlights the lower current region to show the drift in the current.
Only discontinuities up to \SI{2}{pA} are thus shown, however, these discontinuities can go as high as \SI{2}{nA} in the measurement.
The background current can be attributed to an averaged pyroelectric charge generation over the sample.
Since the pyroelectric charge generation is proportional to the time derivative of the temperature, the current should change direction between the cooling process and heating process.
The temperature of the sample is shown in Fig.~\ref{fig:PicData}~(c), showing that the temperature is changed from \SI{280}{K} to \SI{25}{K} and back with a constant rate of temperature change.
After the sample reaches \SI{25}{K}, it remains at this temperature for \SI{43}{min}.
    
\begin{figure}[ht]
    \centering
    \includegraphics{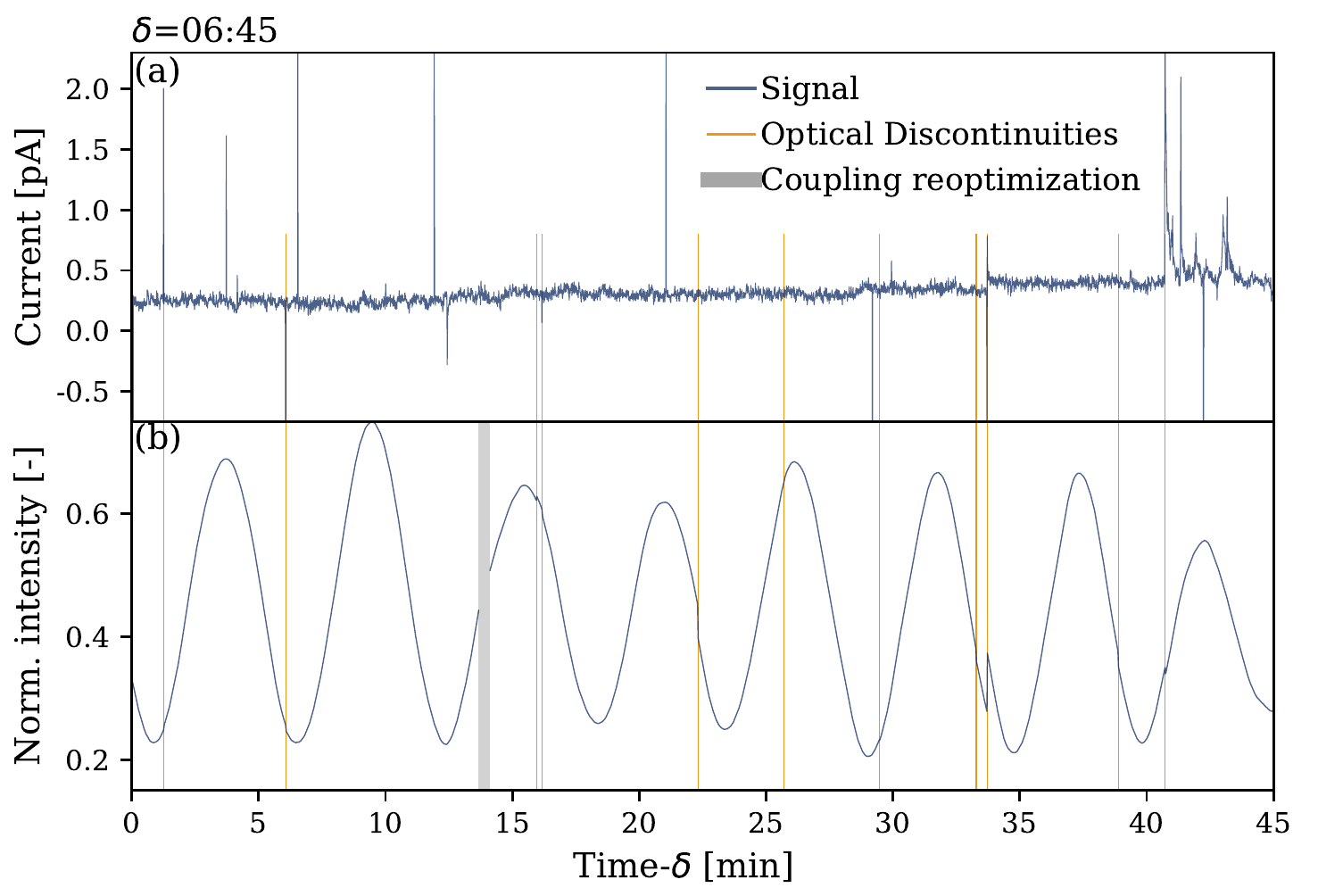}
    \caption{Cutout of the data set in Fig.~\ref{fig:PicData} starting at \SI{127}{K} during the warming process.
    (a) The electrical current measured by the current meter. The spikes are electrical discontinuities and indicate the recombination of charges.
    (b) Normalized optical intensity $\tilde{I}_{\textrm{1}}$.
    The gray shaded area marks the time interval where the optical power was reoptimized, and the orange vertical lines mark all optical discontinuities extracted by the algorithm.
    }
    \label{fig:PicCutout}
\end{figure}

We analyze the correlation between the electrical and optical discontinuities, which results in three different types of possibilities:
    \begin{itemize}
        \item Purely optical discontinuity;
        \item Purely electrical discontinuity;
        \item Correlated discontinuities.
    \end{itemize}

All three possibilities can be found in Fig.~\ref{fig:PicCutout}, showing a cutout of the dataset starting at 
$\delta=06\mathpunct{:}45\mathrm{h}$ and has a length of \SI{45}{min}, spanning the temperature range from \SI{127}{K} to \SI{172}{K}.
The yellow shaded area (at \SI{15}{minutes}) marks a time interval where the optical power is reoptimized, and the vertical dashed lines mark the times where an optical discontinuity is extracted.
Electrical discontinuities are visible from the signal itself as spikes in the electrical signal, and multiple datapoints exceed the vertical limit of the figure.
This figure shows clearly that there are individual discontinuities in the electrical signal (e.g.~\SI{12}{min}) and optical signal (e.g.~\SI{22}{min}), as well as correlated discontinuities (e.g.~\SI{33}{min}).
The electrical discontinuities after \SI{40}{minutes} show an exponential decay, indicating relaxation of electrical charges.

\begin{figure}[ht]
    \centering
        \includegraphics{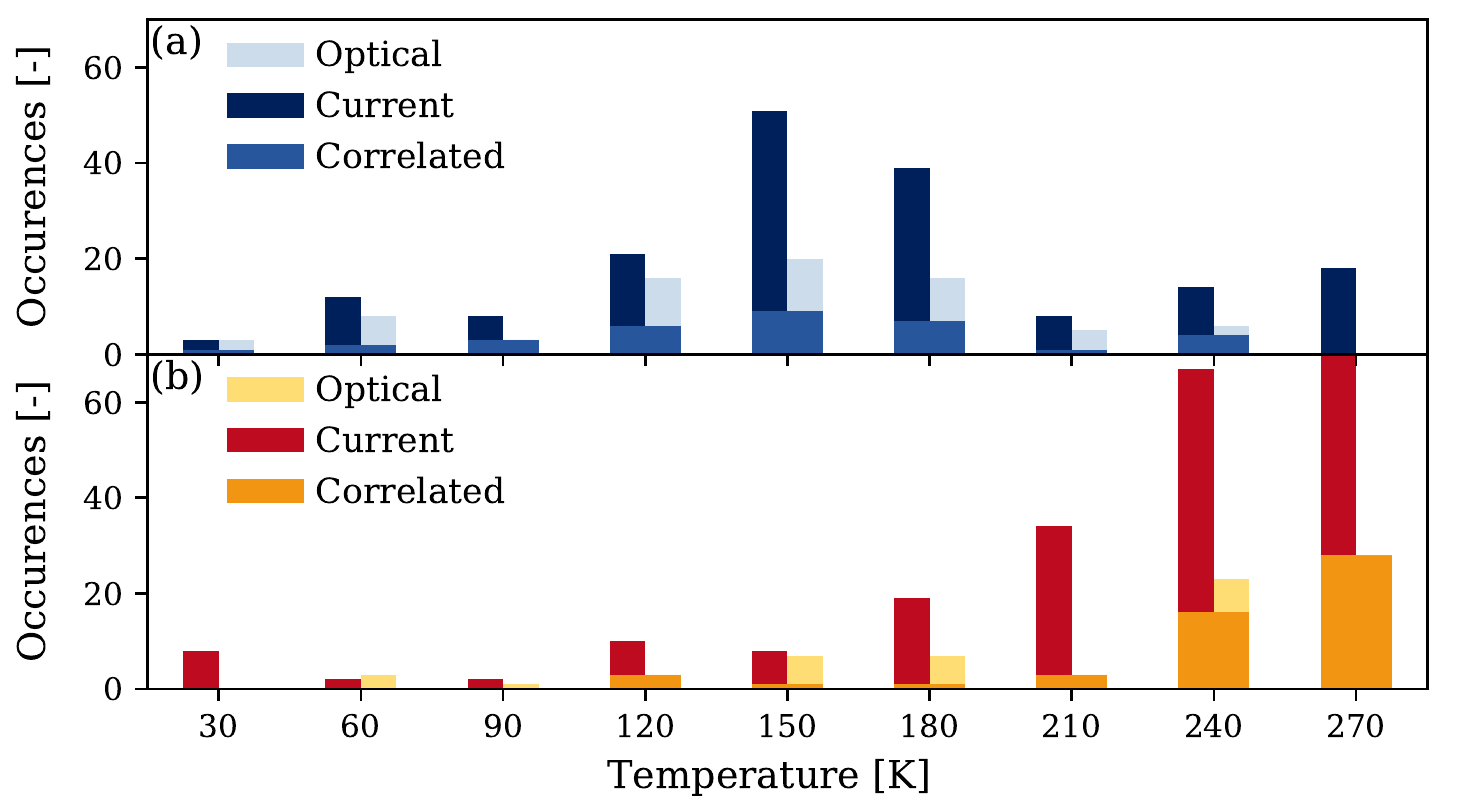}
    \caption{Histogram of the number of discontinuities (occurrences) for the electrical and optical signal, as well as for correlated discontinuities.
    The binning is chosen with bin centres from \SI{30}{K} to \SI{270}{K} with binwidths of \SI{30}{K}, spanning the full temperature range of our measurement.
    (a) shows the number of occurrences during the cooling process, and (b) during the warming process.
    They show different temperature ranges where the number of discontinuities reaches a local maximum.
    In (b) we truncated the number of electrical discharges at \SI{270}{K} reaching \SI{200} occurrences, for an improved visibility.
    However, both show that the number of discontinuities almost disappears below \SI{115}{K}.
    }
    \label{fig:PicHistogram}
\end{figure}

Figure~\ref{fig:PicData} indicates that there are temperature intervals with higher and lower densities of discontinuities.
For this we investigate the occurrence of discontinuities over the different temperature intervals, with a binning width of \SI{30}{K} and bin centres from \SI{30}{K} to \SI{270}{K}.
This range is chosen since these binwidths cover the full temperature range with a significant number of steps. 
In addition, a significant number of occurrences are maintained in most bins.
Figure~\ref{fig:PicHistogram} shows the occurrences of optical and electrical discontinuities, as well as the correlated discontinuities.
Due to the data acquisition and signal analysis, electrical and optical discontinuities within \SI{6}{s} of each other are considered correlated.
Histograms are made for both the cooling process (Fig.~\ref{fig:PicHistogram}~(a)) and the warming process (Fig.~\ref{fig:PicHistogram}~(b)).
During the sample cooling, a clear increase in the number of discontinuities between \SI{195}{K} and \SI{115}{K} is present.
In this range, the electrical occurrences dominate over the occurrences of optical discontinuities.
Below \SI{115}{K} discontinuities are less frequent, indicating a reduced pyroelectric effect resulting in higher stability compared to the temperatures above \SI{115}{K}.
Warming up the sample shows an increase of the number of discontinuities with the temperature starting from \SI{115}{K}.
These two histograms imply that the temperature at which most discontinuities occur, depends on whether the sample is warmed up or cooled down. However, in both cases they almost disappear below \SI{115}{K}.
The reduction of the pyroelectric charge recombination can be explained by the reduced pyroelectric coefficient~\cite{Shaldin2008}, and reduced charge mobility~\cite{Kotitz1990,Akhmadullin1998}, at cryogenic temperatures.

We assume that almost no unbound charge carriers are present in the sample when we start the experiment at room temperature. 
During the cooling process, pyroelectric charges are generated and accumulate within the material domain until the charge density reaches a threshold and the charges recombine.
Lower temperatures should result in a lower probability generating pyroelectric charges due to the reduced pyroelectric coefficient.
In addition, the charge mobility is reduced, increasing the threshold for the required discharge to warrant a recombination.
These two effects cause a reduction of charge recombinations at cryogenic temperatures, which can be seen from the reduced number of electrical discontinuities below \SI{115}{K}.
This results in a reduced number of discharges while the sample is at a stable cryogenic temperature.
When the sample is warmed up again, the charge mobility increases, and more pyroelectric charges are generated again which results in electrical discharges~\cite{Kotitz1990, Shaldin2008}. 
As a result, the probability of charge relaxation will increase with temperature, resulting in an ever in creasing number of electrical discontinuities at our highest measured temperature.

\begin{figure}[ht]
    \centering
        \includegraphics{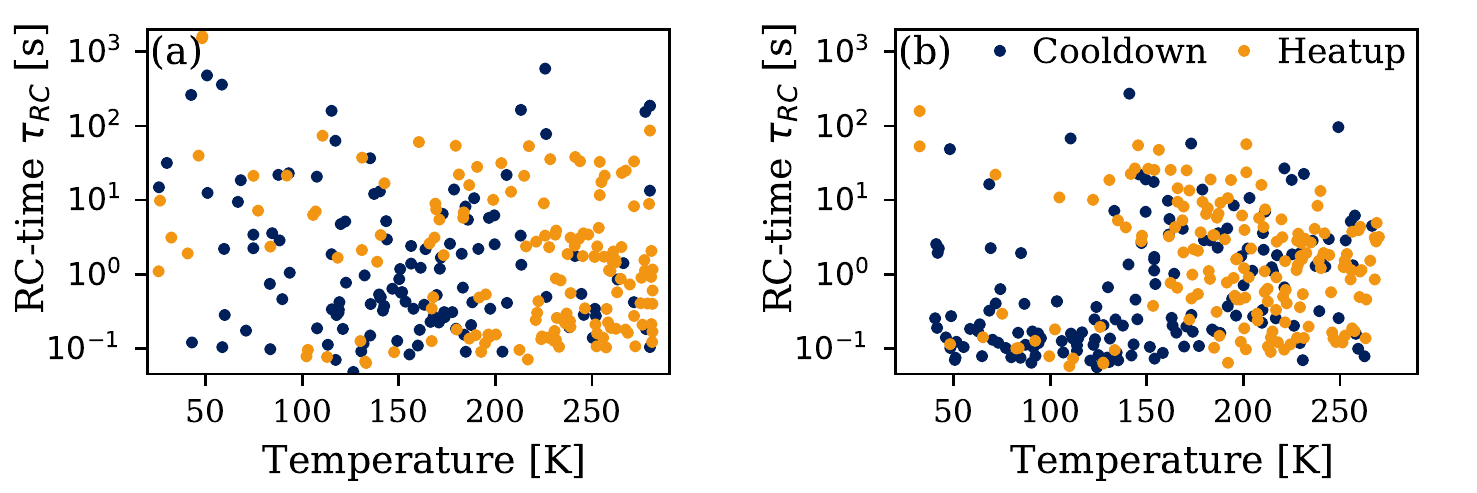}
    \caption{
    Decay times of the identified current fluctuations for the measurement (a) with light, and (b) without light coupled through the waveguide.
    Current fluctuations with only a single data point are excluded since no decay rate can be fitted to these.}
    \label{fig:PicRC}
\end{figure}

Our method allows us to extract current fluctuations in the sample. 
Furthermore, we analyse these discontinuities upon their decaying dynamics in the charge recombination. The recombination of the charges depends on the effective resistance and capacitance from the point of origin to the recombination. 
As a result, we can connect an RC-time ($\tau_{RC}$) to every discharge by fitting an exponential decay~\cite{Korotky1996}.
Fluctuations with only a single point decay are excluded since they are too fast to be discriminated with our set sampling rate.
The RC-time of all other decays is shown in Fig.~\ref{fig:PicRC}~(a), where it is plotted versus the temperature, and separated by cool down (blue dots) and heat up (orange dots).
The extracted decays show that the decay rates are independent on the temperature and vary over multiple orders of magnitude. 
Additionally, the decay rates are independent of direction in the temperature change, however, the decay density depends on the temperature as was shown in Fig.~\ref{fig:PicHistogram}.
This is expected since carriers are generated at various points in the material domain and redistribute in different directions, resulting in different RC-times.

Unbound charges can also be generated by the transmitted light, the so called photorefractive effect~\cite{Boyd2007}. 
These unbound carriers could be another source to induce electric fields in the vicinity of the lithium niobate waveguide. 
To investigate this influence, we additionally cooled down and heated up the sample without light going through the waveguide and with the same temperature rates (\SI{1}{K/min}), while we measured the current through the electrodes. 
In the thermal transition, fluctuations in the current are visible similar to the measurements with light through the waveguide. 
The number of occurrences and the related RC-times are similar to the optical measurement, as it can be seen in Fig.~\ref{fig:PicRC}~(b). 
This data hints that the measurement without light shows more fast decays at cryogenic temperature, thus less discontinuities when light is coupled through the waveguide.
However, due to the sparsity of the data, no hard conclusion can be drawn.    

\section{Conclusion}
Lithium niobate has proven to be a valuable material system for integrated quantum optics, expanding into cryogenic temperature regimes~\cite{Thiele2020,Hopker2019,hopker2021integrated,Lomonte2020,Bartnick2021,Thiele2022a,Lange2022}. 
In this photonic platform, cryogenic frequency conversion sources~\cite{Bartnick2021,Lange2022} and electro-optic modulators~\cite{Thiele2020,Thiele2022a} were implemented, revealing optical changes in the operation characteristics. 
Furthermore, the integration of superconducting single photon detectors poses challenges due to the destructive nature of pyroelectric charges during the cooling process~\cite{hopker2021integrated}.

To address this issue and create a more reliable platform for integrated quantum photonics, we investigated the optical changes and pyroelectric discharges in lithium niobate. 
We utilize the Sénarmont method to measure changes in waveguide birefringence and correlate these changes with acquired pyroelectric discharges. 
The occurrence of electronic and optical perturbations show a correlation during the entire temperature range.

During the cooling process, the occurrences of pyroelectric discharges and optical changes increase until a temperature of approximately \SI{150}{K} is reached, afterwards they reduce significantly and below \SI{100}{K} nearly vanishes.
At stable cryogenic temperatures, minimal optical or current perturbations are observed. 
During the heating process, the rate of occurrences is constant until \SI{100}{K} and increases for higher temperatures.

To further investigate the phenomenon, we conducted control measurements without any input light. 
The comparison between the control and Sénarmont measurements shows no significant variations in the generated pyroelectric charges, excluding the significant influence of photo-refraction. 
This confirms that pyroelectric charge generation during a temperature transition is the predominant effect.

The decay rates of the electrical discharges were found to span several orders of magnitude, regardless of the temperature. 
This suggests that there is no specific capacitive geometry within the sample where all the discharges occur.
Future investigations can explore the impact of different types and designs of electro-optic modulators, as well as the effect of the temperature rate during thermal transitions.
By understanding these effects, introduction of partial conductive layers on the lithium niobate structures to minimize the influence of pyroelectric charges can be investigated.

By studying these optical changes and pyroelectric discharges in lithium niobate, we aim to enable the integration of superconducting single photon detectors with modulators and frequency conversion sources. 
This integration is crucial for achieving a scalable photonic platform in the field of integrated quantum photonics.

\section{Acknowledgements}
This work was supported by the Bundesministerium für Bildung und Forschung (Grant No. 13N14911) and the Deutsche
Forschungsgemeinschaft (231447078–TRR 142). 

\section{References}

\bibliographystyle{ieeetr}
\bibliography{bib2}
\end{document}